\documentclass[12pt,preprint]{aastex}

\shorttitle{Sulfur chemistry and isotopic ratios in NGC\,253}
\shortauthors{Mart\'{\i}n et al.}

\usepackage{graphicx}
\begin{document}
\title{Sulfur chemistry and isotopic ratios in the starburst galaxy NGC\,253}

\shorttitle{Sulfur chemistry in NGC\,253}

\author{S. Mart\'{\i}n}
\affil{Instituto de Radioastronom\'{\i}a Milim\'etrica (IRAM), Avda. Divina Pastora 7, Local 20, E-18012 Granada, Spain}
\author{J. Mart\'{\i}n-Pintado}
\affil{Departamento de Astrofis\'{\i}ca Molecular e Infrarroja, Instituto de Estructura de la Materia, CSIC, Serrano 121, E-28006 Madrid, Spain}
\author{R. Mauersberger}
\affil{Instituto de Radioastronom\'{\i}a Milim\'etrica (IRAM), Avda. Divina Pastora 7, Local 20, E-18012 Granada, Spain}
\author{C. Henkel}
\affil{Max-Planck-Institut f\"ur Radioastronomie, Auf dem H\"ugel 69, D-53121 Bonn, Germany}
\author{S. Garc\'{\i}a-Burillo}
\affil{Observatorio de Madrid, Alfonso XII, 3, 28014 Madrid, Spain}

\begin{abstract}
Based on observations of the most abundant sulfur-bearing molecules (H$_2$S, CS, NS, SO, H$_2$CS, OCS, and SO$_2$) carried out with
the IRAM 30m telescope
\footnote{Based on observations carried out with the IRAM 30m telescope.
IRAM is supported by INSU/CNRS (France), MPG (Germany) and IGN (Spain).} and SEST
\footnote{Based on observations with the Swedish/ESO submillimetre telescope (SEST) at the European Southern Observatory, 
Chile. ESO N$^{\rm o}$70.B-0022.},
we present the first analysis of the sulfur chemistry of an extragalactic source, the nuclear region of the
starburst galaxy NGC\,253.
This is the first time that H$_2$S and, tentatively, H$_2$CS are
detected towards the nucleus of a starburst galaxy. 
Source averaged fractional abundances of these molecules are a few 10$^{-9}$,
except for CS and OCS which are more abundant (10$^{-8}$).
Sulfur isotopic ratios, $^{32}$S/$^{34}$S$\sim$8$\pm 2$ and $^{34}$S/$^{33}$S$>$9, are measured through observations of $^{13}$CS, C$^{34}$S,
and C$^{33}$S.
A comparison with the observed relative abundances towards different prototypical Galactic sources suggests that the chemical composition
of NGC\,253 is similar to that found towards the molecular clouds complexes like Sgr\,B2 in the nuclear region of the Milky Way.
The large overabundance of OCS compared to the predictions of time-dependent sulfur chemistry models supports the idea
that OCS is likely injected into the gas phase from the grain mantles by low velocity shocks.

\end{abstract}


\section{Introduction}
The formation and evolution of stars strongly affects the chemistry of their molecular environment.
In particular, sulfur-bearing molecules show enhanced abundances in the hot molecular cores, the condensations associated with massive protostars.
The most accepted idea is that during the collapse of a star forming cloud, a large fraction of the available sulfur 
freezes out by accretion onto grain mantles and hydrogenates in the form of H$_2$S.
The high temperatures and turbulent motions in these objects produced by the newly formed stars induce the evaporation and disruption
of grain mantles, releasing the H$_2$S into the gas phase.
After evaporation, this molecule is believed to drive a fast high temperature gas-phase chemistry leading first to the formation of SO and SO$_2$ and
subsequently of other sulfur-bearing molecules such as CS and H$_2$CS \citep{Charnley97,Hatchell98}.
It is still uncertain whether the release of H$_2$S is caused exclusively by thermal evaporation or whether
shocks play a role in the disruption of the grain mantles \citep{Hatchell02}.
In low mass star-forming regions, where the temperatures
are not high enough to evaporate the grain mantles,
there is evidence that shocks may be responsible for injecting the H$_2$S from ices into the gas-phase \citep{Buckle03}.
Sulfur-bearing molecules are considered to be an important tool for studying the presence of shocks within
 massive star-forming regions.
Time-dependent chemical models have roughly succeeded to reproduce the observed abundances of the most common sulfur-bearing species both in high and
low mass star-forming regions \citep{Hatchell98,Buckle03}.

In starburst galaxies like NGC\,253, only the most abundant sulfur-bearing molecules have been detected so far, namely CS, OCS, SO
\citep{Henkel85,Mauers95,Petu92},
and recently also SO$_2$ and NS \citep{Martin03}.
Extragalactic H$_2$S has only been detected toward the Large Magellanic Cloud, where H$_2$CS is also tentatively detected
\citep{Heikkila99}.
In this paper we present
the first detection of H$_2$S and the tentative detection of
H$_2$CS in NGC\,253, allowing us to investigate for the first time the sulfur chemistry in its nuclear region
using a set of the most abundant sulfur-bearing species.

\section{Observations}
Most of the observations of the molecular species listed in Table~\ref{tab:fits} were carried out with the IRAM 30\,m telescope on Pico
Veleta (Spain). The measurements were made in symmetrical wobbler
switching mode with a frequency of 0.5\,Hz and a beamthrow of $4'$ in azimuth.
As spectrometers, we used two $256\times4$\,MHz filterbanks for the transitions at 2\,mm and
two $512\times1$\,MHz filterbanks for those at 3\,mm.
The $J=2-1$ and $J=5-4$ transitions of CS were measured with
the 15\,m Swedish-ESO Submillimetre Telescope (SEST) at La Silla (Chile).
Observations were done in wide dual beam switch mode, with $11.8'$ beam throw in azimuth.
Acousto Optical Spectrometers were used as backends with a resolution of 1.4\,MHz.
Beam sizes ranged from $29''$ (at 85\,GHz) to $12''$ (at 196\,GHz) for the 30m telescope and from $51''$ (at 98\,GHz) to $21''$ (at 244\,GHz) for the SEST.
In both telescopes, receivers were tuned to single side band with an image band rejection larger than 10dB.
The spectra were calibrated with a dual load system.

\begin{figure}[ht]
	\begin{minipage}[b]{0.3\textwidth}
		\includegraphics[width=\linewidth]{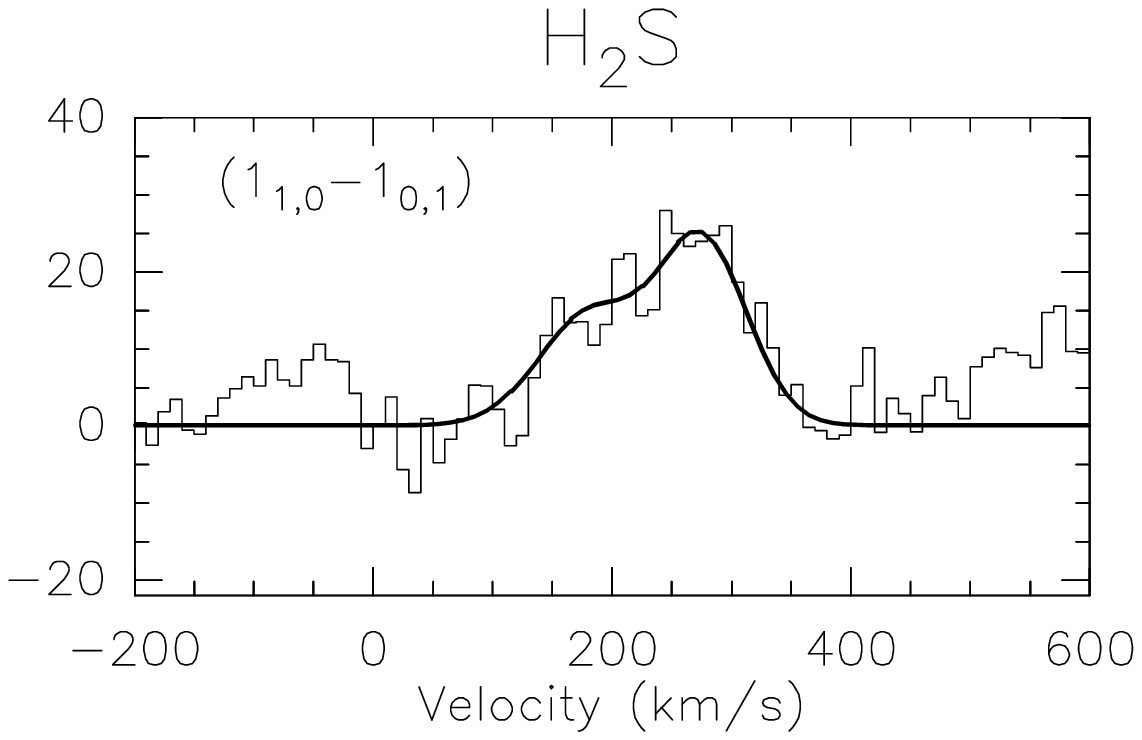}
		\\[20 pt]
		\includegraphics[width=\linewidth]{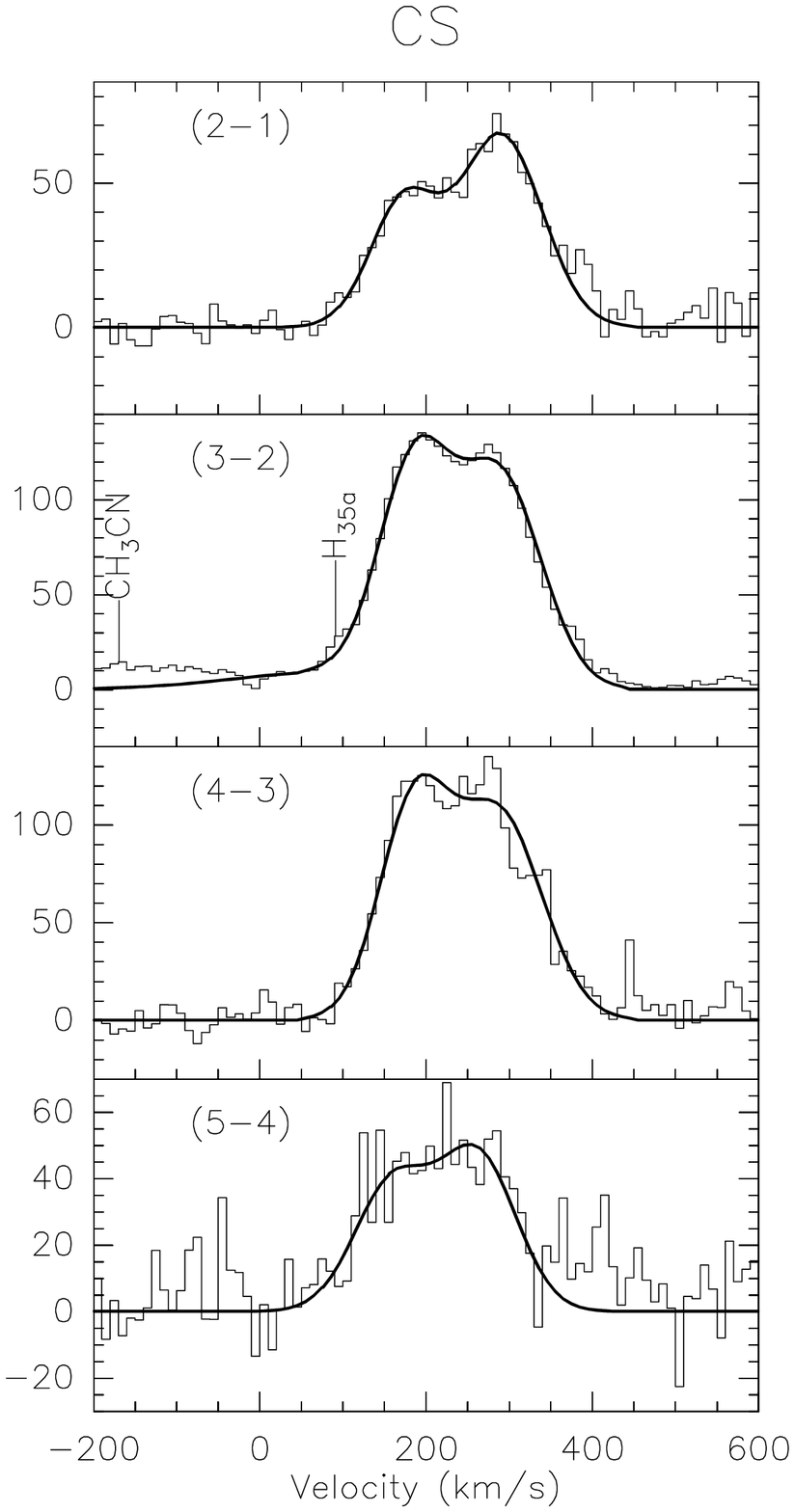}
	\end{minipage}%
	\hspace{5 pt}
	\begin{minipage}[b]{0.3\textwidth}
		\includegraphics[width=\linewidth]{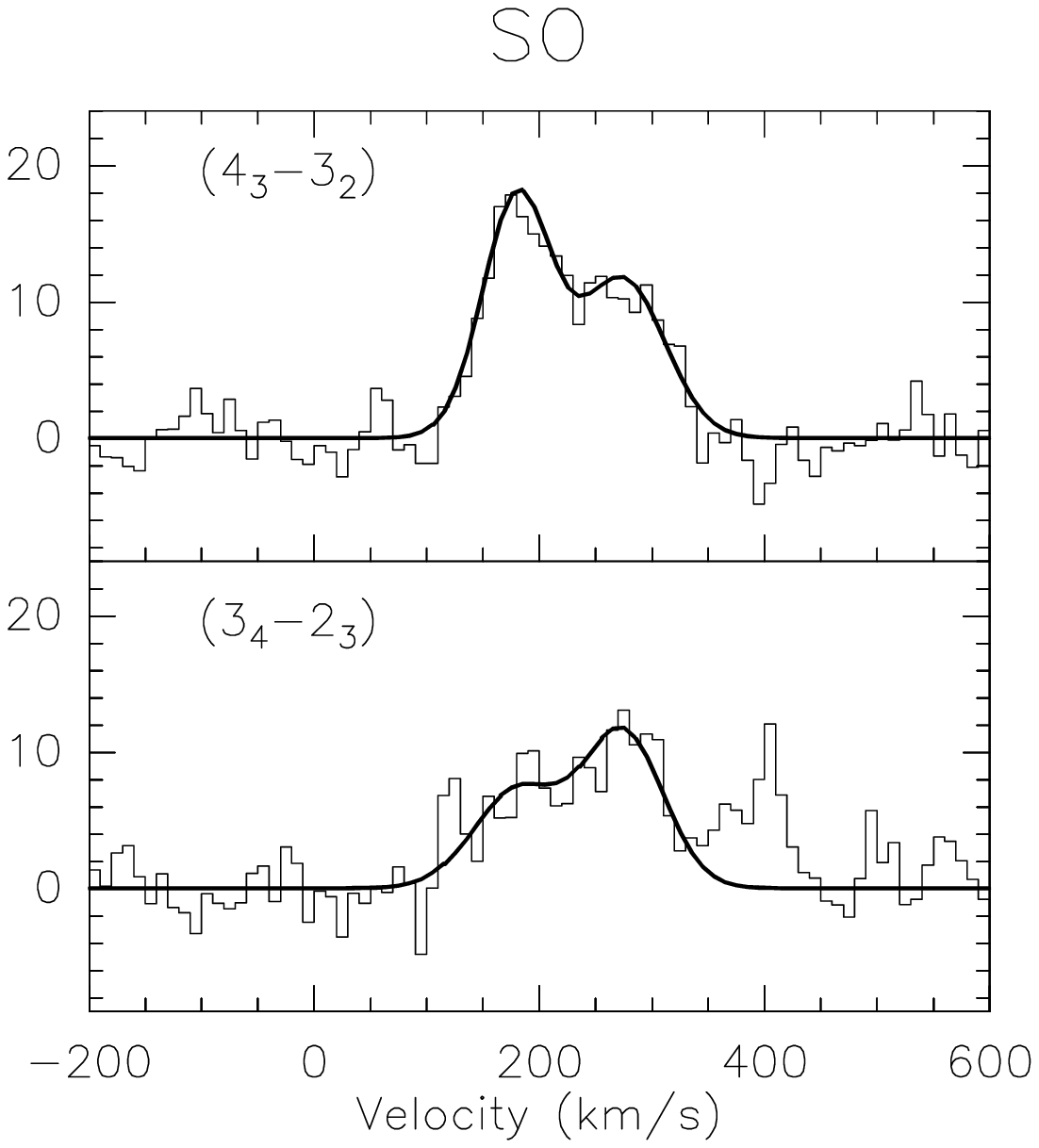}
		\\[10 pt]
		\includegraphics[width=\linewidth]{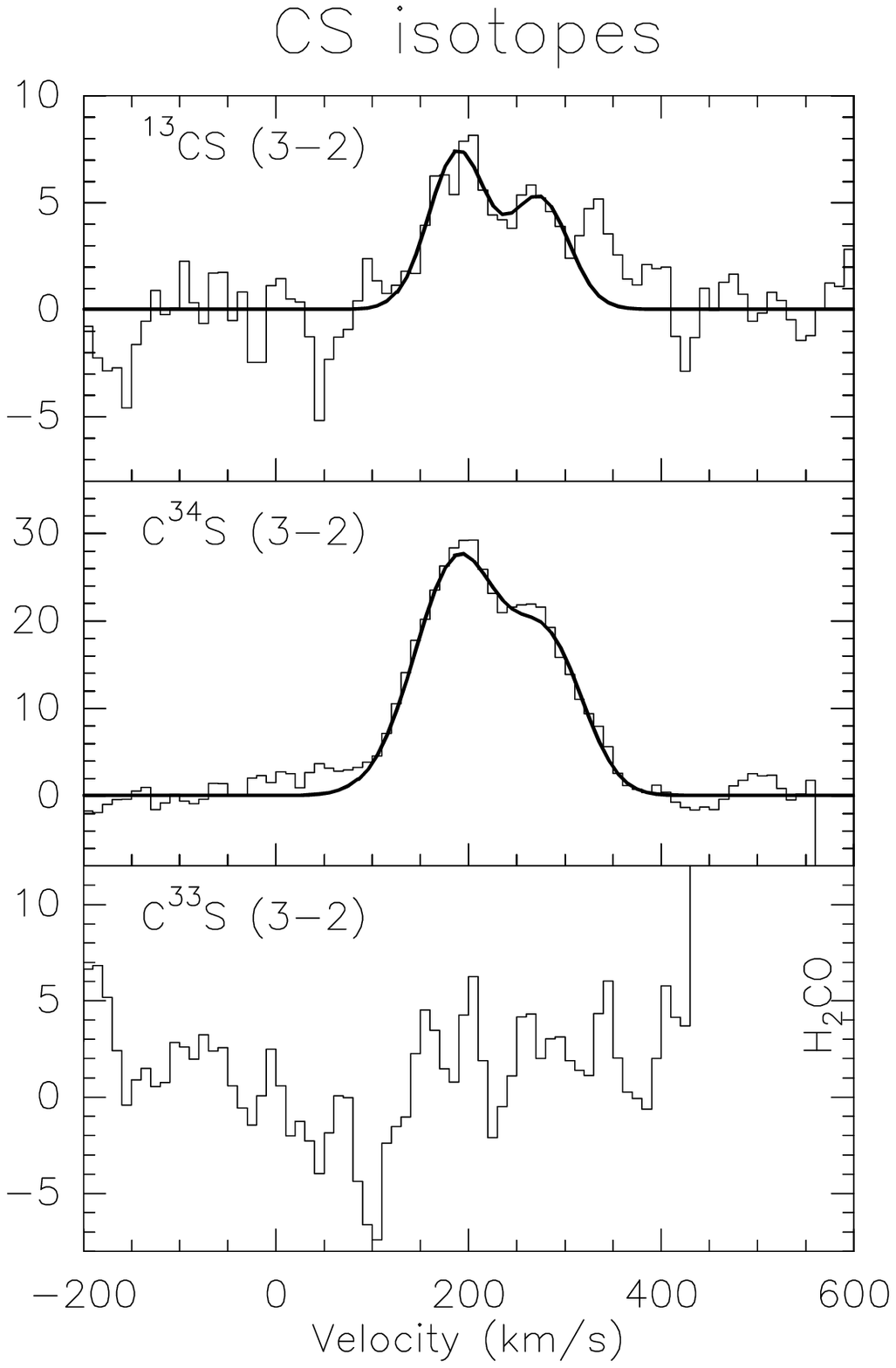}
	\end{minipage}
	\begin{minipage}[b]{0.3\textwidth}
			\includegraphics[width=\linewidth]{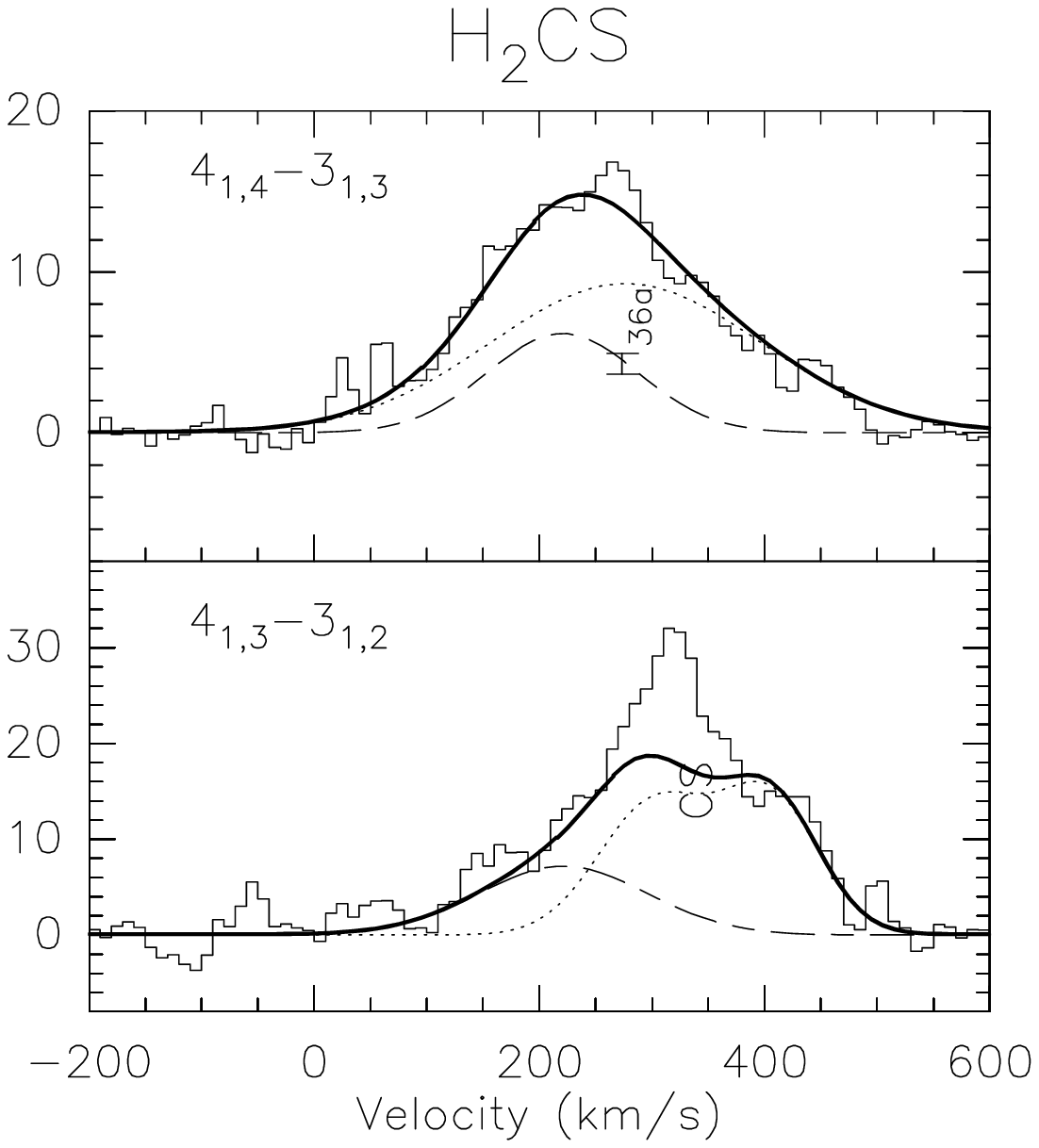}
			\\[5 pt]
			\includegraphics[width=\linewidth]{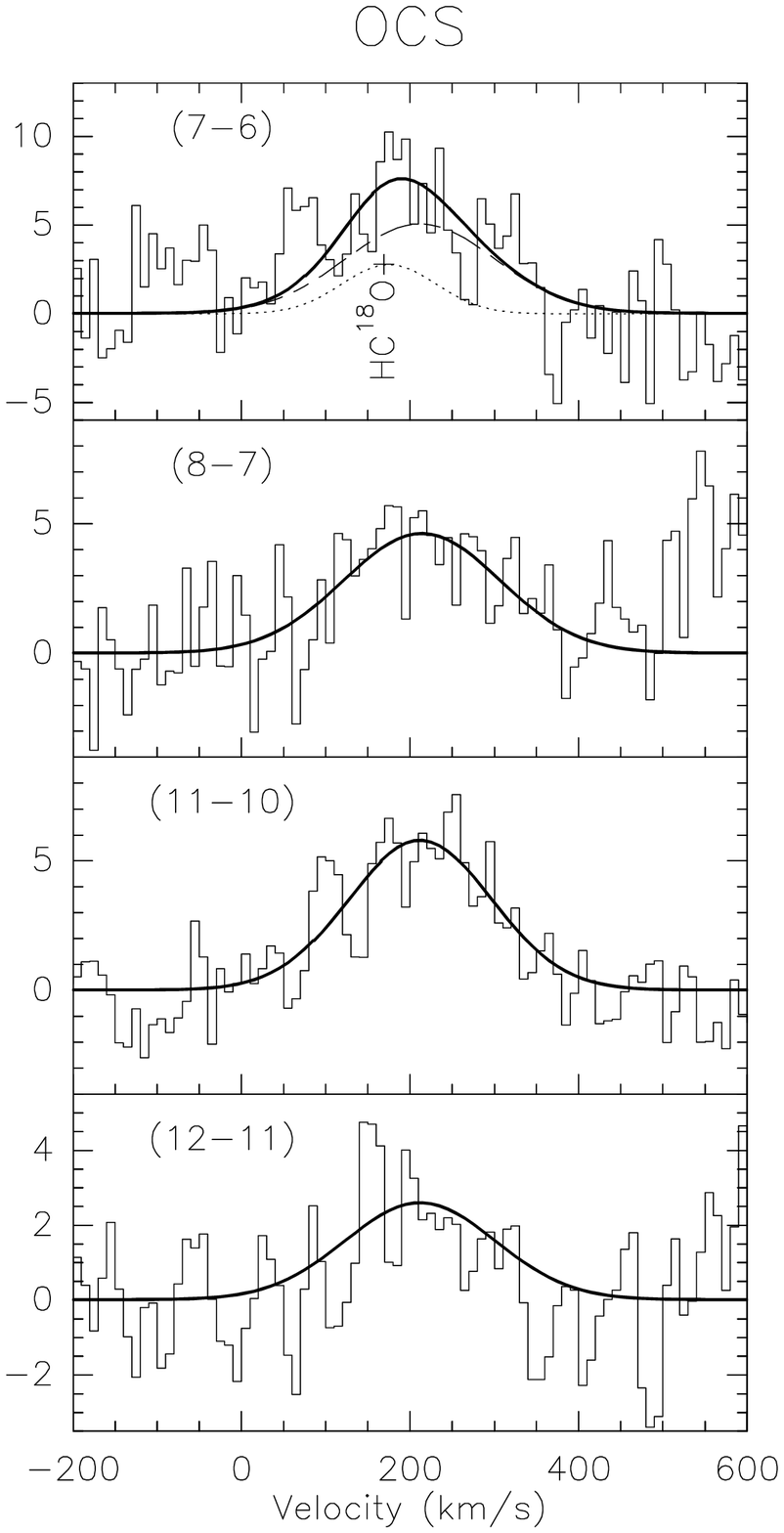}
	\end{minipage}
  \figcaption {Gaussian fits to the detected transitions of H$_2$S, CS, $^{13}$CS, C$^{34}$S, SO, H$_2$CS, and OCS towards the nuclear
  region of NGC\,253 ($\alpha_{J2000}=00^{\rm h}47^{\rm m}33\fs4 ,\delta_{J2000}=-25\arcdeg17\arcmin23\arcsec$)
  The velocity resolution has been smoothed to 10\,km\,s$^{-1}$. Intensities are given in $T_{\rm MB}$ (mK).
  The figure for H$_2$CS shows in dashed lines the fits to the H$_2$CS transitions, in dotted lines the assumed H36$\alpha$ and CS profiles
  and in thick line the total fit. Similarly, the figure for the OCS $J=7-6$ transition shows the calculated HC$^{18}$O$^{+}$ in dotted line.
  \label{fig:transitions}}
\end{figure}

Fig.~\ref{fig:transitions} shows the observed line profiles of CS and two of its isotopic substitutions ($^{13}$CS and C$^{34}$S),
H$_2$S, SO, H$_2$CS, and OCS, as well as the corresponding Gaussian fits.

\section{Results}
\label{sect:results}
For most of the observed transitions, we identify two velocity components, which arise from molecular cloud complexes
at two opposing sides of the nucleus \citep{Mauers96}, located at $\sim 5''$ NE (the 180\,km\,s$^{-1}$ component) and
$\sim 5 ''$ SW (the 280\,km\,s$^{-1}$ component) as revealed by interferometric maps \citep{Peng96,Burillo00}.

The derived parameters from the Gaussian fits to the observed profiles are shown in Table~\ref{tab:fits}.
While most of the transitions of the main isotopic species of CS have already been reported \citep{Mauers89,Henkel93},
the improved signal-to-noise ratios of our data allow us for the first time to separate the two main velocity
components in the line profiles.
Unfortunately, the 2\,mm lines of H$_2$CS appear to be blended with other lines.
The $4_{1,4}-3_{1,3}$ transition is blended with the H$36\alpha$ recombination line and the
$4_{1,3}-3_{1,2}$ transition is contaminated by a contribution of the CS $J=3-2$ line from the image band.
In order to perform the Gaussian fits to the H$_2$CS lines, we have respectively subtracted a Gaussian profile similar to that
observed for the H$34\alpha$ line measured at 160\,GHz and the CS $J=3-2$ emission measured in the signal band corrected
by the corresponding image gain.
As seen in the lower panel of the H$_2$CS lines in Fig.~\ref{fig:transitions}, the CS and the H$_2$CS lines do not fully account
for the observed feature.
The extra emission could be due to the presence of another unidentified line.
In spite of the uncertainties due to the blending, both resulting H$_2$CS profiles
are self-consistent, since they are expected to have similar intensities. Nevertheless, this detection must be
regarded as tentative.
Similarly, the OCS $J=7-6$ transition appears blended with the $J=1-0$ line of HC$^{18}$O$^+$. Additional observation of the HC$^{18}$O$^+$ $J=2-1$
transition at 170\,GHz allowed us to estimate and subtract the contribution of this species to the observed feature,
by assuming local thermodynamic equilibrium (LTE) with an excitation temperature $T_{\rm ex}\sim$12\,K.

\subsection{Molecular abundances and excitation temperatures}
In order to derive the physical properties of the molecular material we need to make an assumption on the total emission extent 
of the molecular emitting region. By smoothing the interferometric CS data from \citet{Peng96}, \citet{Mauers03} derived the dependence 
of the observed line intensity on the beam size. This dependence indicates an equivalent source size of $\sim20''$ for the
CS emitting region. The interferometer maps of CS only miss 36\% of the flux \citep{Peng96} suggesting that the extent of the CS emission is not
much larger than this estimate. As we have no additional information on the extent of the emission of the other rarer species, we will 
assume them to be similar to that of CS. This assumption is supported by the comparison of the high angular resolution maps of the CS emission
with those of species with a fairly different chemistry such as SiO, H$^{13}$CO$^+$ \citep{Burillo00} and NH$_3$ (M. Lebron, private communication), 
where the morphology of the emission seems to be similar.

\begin{figure}[ht]
\begin{center}
\includegraphics[width=0.23\linewidth, angle=-90]{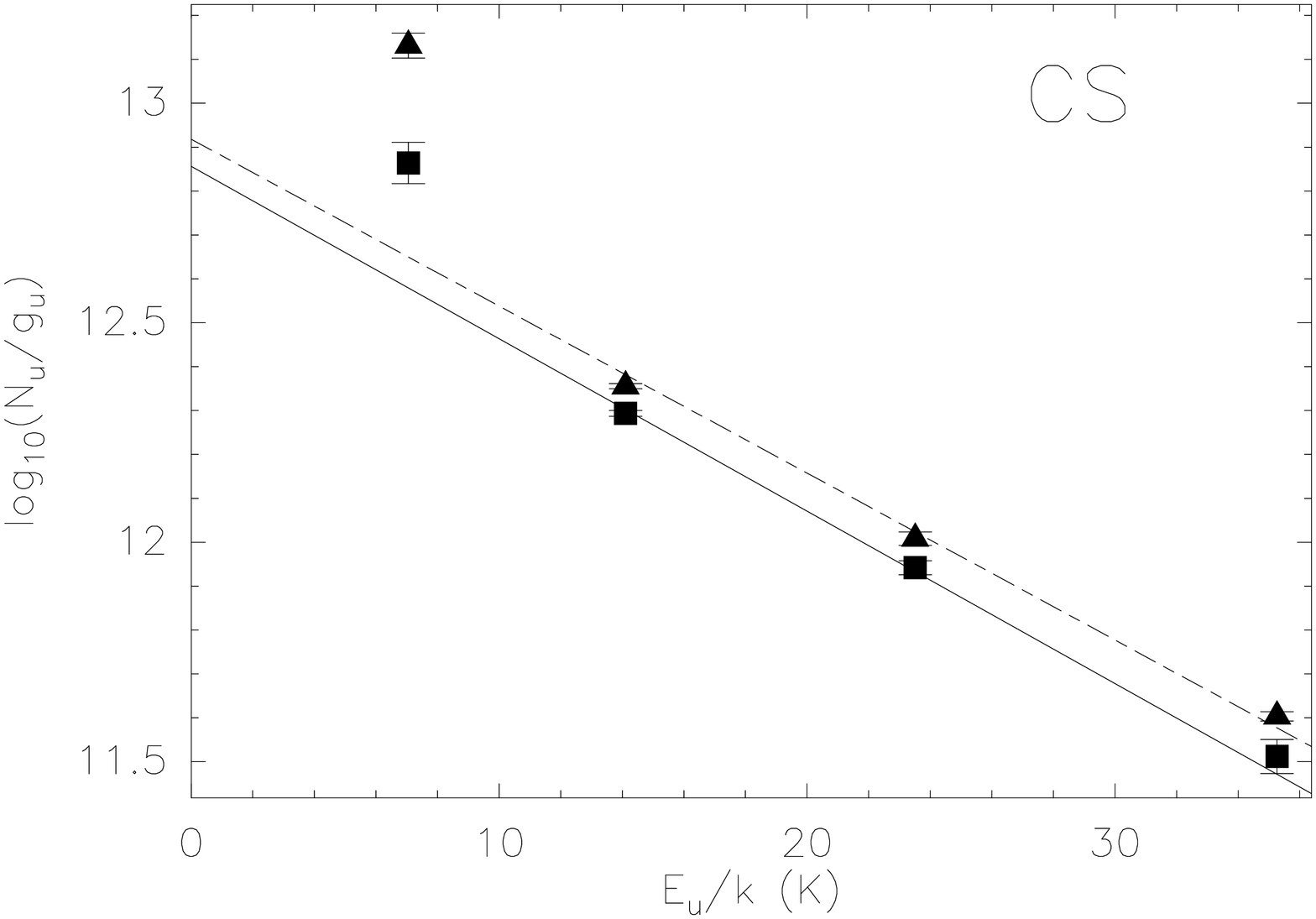}
\includegraphics[width=0.23\linewidth, angle=-90]{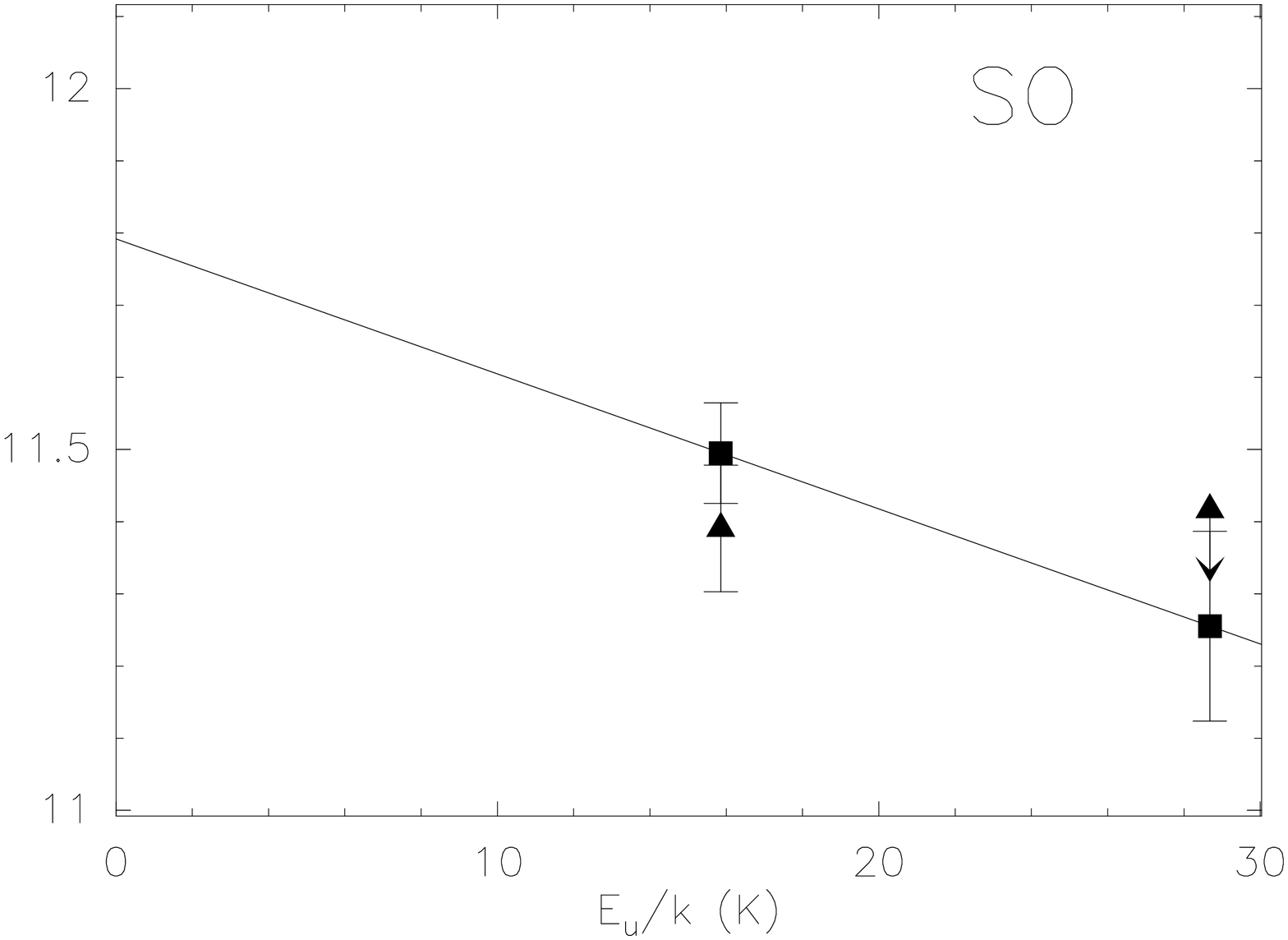}
\includegraphics[width=0.23\linewidth, angle=-90]{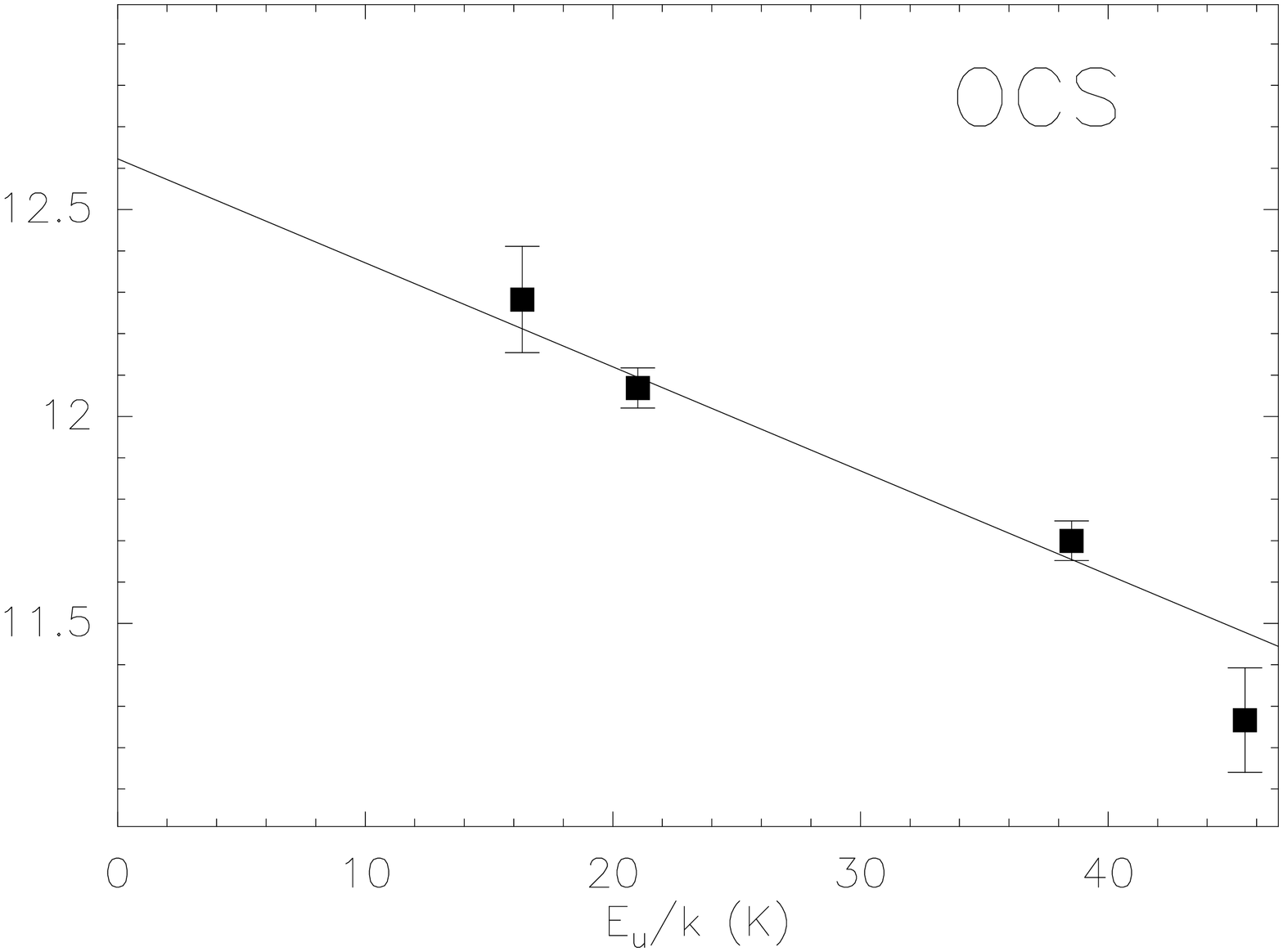}
\figcaption {Population diagrams for the CS, SO, and OCS molecules. 180\,km\,s$^{-1}$ component is represented by squares and continuous
lines, the 280\,km\,s$^{-1}$ component is shown with triangles and dashed lines.
\label{fig:diagrams}}
\end{center}
\end{figure}

We have derived source averaged column densities by correcting the measured line intensities for the estimated source size of $20''$
assuming optically thin emission. The corresponding
population diagrams for CS, SO, and OCS are shown in Fig.~\ref{fig:diagrams}.
For CS we show the population diagrams for the two velocity components at 180 and 280\,km\,s$^{-1}$ and we derive a
rotational temperature $T_{\rm rot}\sim10\,$K for both components. In these diagrams, the population in the $J=2$ level, clearly above the
linear fit, shows evidence for the presence of a lower excitation temperature component with $T_{\rm rot}\sim 5$\,K which
can be explained by the much larger region observed by the $51''$ beam of the SEST.
As derived below, the CS emission is moderately optically thick.
The effect of the opacity on the population diagram would result in a slight increase of the derived excitation temperatures.
Thus we would get temperatures of $\sim 13$\,K and $\sim 8$\,K for the two components.
To derive the CS column density in Table~\ref{tab:MolecDensity} we have used the optically thin emission of one of its isotopes, $^{13}$CS.
For OCS
we derive a $T_{\rm rot}=16$\,K.
Our estimate of the OCS column density in Table~\ref{tab:MolecDensity} is in agreement with that derived by \citet{Mauers95}
when the differences in the assumed source size are taken into account.
For the 180\,km\,s$^{-1}$ component of SO we obtain a $T_{\rm rot}=23$\,K.
As seen in Fig.~\ref{fig:diagrams}, the  280\,km\,s$^{-1}$ component of the
SO $3_4-2_3$ transition (E$_{\rm u}/k\sim29$\,K) has a higher intensity than expected if we assume, from the CS results,
both components to have a similar rotational temperature (Fig.~\ref{fig:diagrams}).
This may be due to the contamimation of this component by the emission feature observed at a velocity of $\sim400$
km\,s$^{-1}$ which we tentatively identify as the $8_{1,8}-7_{1,7}$ transition of NH$_2$CN (Cyanamide).
The measured integrated intensity has been then considered as an upper limit as shown in Fig.~\ref{fig:diagrams}.
We have therefore assumed a similar excitation temperature of 23\,K for both velocity components to
compute the column density of the 280\,km\,s$^{-1}$ component of SO.
Since we have only one transition for H$_2$S and the two observed lines of H$_2$CS have the same upper level energy,
we need to make an assumption on the excitation temperature of these species.
As the Einstein coefficients of these transitions are similar to those of CS and SO$_2$,
we will take an excitation temperature of 12\,K for both species according to that derived from CS in this
work, and from SO$_2$ by \citet{Martin03}.

Table~\ref{tab:MolecDensity} shows the derived source averaged column densities, rotational temperatures and
fractional abundances relative to H$_2$ for all the sulfur bearing molecules detected towards the nucleus of NGC\,253.
The main sources of uncertainty in the derived abundances are those from the assumed extent of the emitting region and the H$_2$ column density,
which are assumed to be the same for all the species.
Assuming the extreme case in which the molecular emission is confined to a much smaller region of $\sim10''$ 
the derived source averaged column densities will be larger by less than a factor 2. 
However, these uncertainties are expected to be even smaller when column density ratios between different species are considered.
A similar argument also applies to the uncertainties in the derived fractional abundances introduced by the assumed H$_2$ column density if,
as already discussed, the spatial distribution among different species does not vary substantially.
Since we will focus our discussion on the molecular abundance ratios between different species, these uncertainties
will not affect our conclusions. 

Given that the $J=3-2$ transition of CS,
$^{13}$CS, and C$^{34}$S were observed with the same telescope
we can compare the observed $^{12}$CS to $^{13}$CS line intensity ratio of $21\pm3$ with the $^{12}$C/$^{13}$C
ratio of $40\pm10$ estimated by \citet{Henkel93}.
Assuming that the $^{12}$CS/$^{13}$CS intensity ratio reflects the $^{12}$C/$^{13}$C abundance ratio, we derive an optical depth for the
$^{12}$CS\,$J=3-2$ transition of $\sim1.4$.
Therefore from the measured $^{13}$CS/C$^{34}$S line intensities and the
$^{12}$C/$^{13}$C ratio mentioned above we derive a $^{32}$S/$^{34}$S abundance ratio of $8\pm2$.
From our non detection of the C$^{33}$S line, it is possible to give a $3\sigma$ lower limit to the $^{34}$S/$^{33}$S ratio of $9$.

\begin{figure}[ht]
\begin{center}
\includegraphics[width=0.5\linewidth,angle=-90]{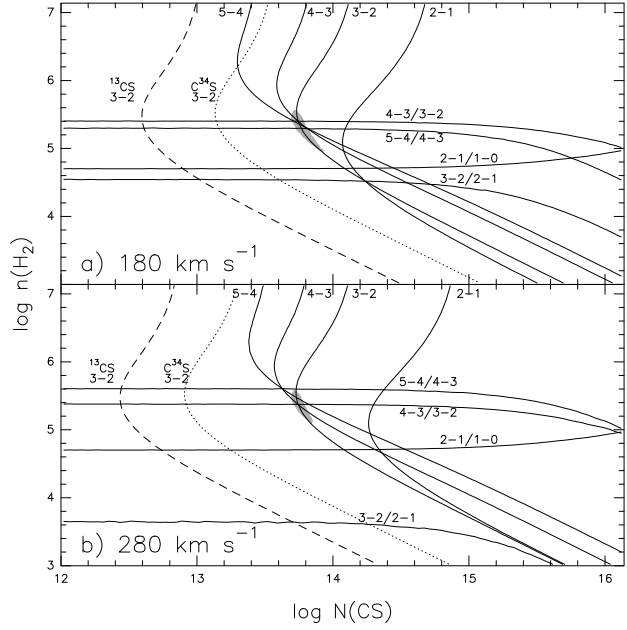}
\figcaption{LVG model predictions for the observed CS transitions and for the $J=3-2$ line of its isotopomers $^{13}$CS and C$^{34}$S at a $T_{\rm kin}$=100\,K
as a function of the H$_2$ density and the CS column density.
Upper and lower pannel correspond to the 180 and 280 \,km\,s$^{-1}$ velocity components respectively.
Lines show the measured line intensities and intensity ratios. The 2--1/1--0 ratio is from \citet{Paglione}.
Best fit to the line intensities of CS that minimizes the reduced $\chi ^2$ function is shown in grey and corresponds to the
probability of 99.5\% of enclosing the true parameters.
\label{fig:CSlvg}}
\end{center}
\end{figure}

The line intensities corrected for beam dilution of the four observed transitions of CS and the $J=3-2$ lines of $^{13}$CS and C$^{34}$S
have been compared with the results of model calculations for the excitation of CS in which the Large Velocity Gradient approximation
has been used.
A linewidth of 100\,km\,s$^{-1}$ and a kinetic temperature $T_{\rm kin}$=100\,K
(similar to the rotational temperature of 100--142\,K derived from NH$_3$ measurements by \citet{Mauers03})
have been considered in these non-LTE calculations.
Fig.~\ref{fig:CSlvg} shows the results of this model for the observed line intensities and line ratios. 
The grey region corresponds to the best fit to the observed CS main isotope line intensities, calculated by minimizing the reduced $\chi^2$ 
function as described by \citet{Nummelin}.
The best fit gives a CS column density
N(CS)=$5.7\times10^{13}$cm$^{-2}$ and a molecular hydrogen density n(H$_2$)=$2.2\times10^{5}$cm$^{-3}$ for the velocity component at
180\,km\,s$^{-1}$ and N(CS)=$5.7\times10^{13}$cm$^{-2}$ and n(H$_2$)=$2.4\times10^{5}$cm$^{-3}$ for that at 280\,km\,s$^{-1}$.
These column densities are a factor of two lower than those in Table~\ref{tab:MolecDensity} derived from the isotopomer $^{13}$CS assuming LTE conditions.
The H$_2$ density derived from this analysis depends on the assumed kinetic temperature. Therefore, a $T_{\rm kin}$ of 50\,K or 25\,K would
lead to H$_2$ densities a factor of 2 and 4 higher respectively. The choice of $T_{\rm kin}$ has, however, almost no effect on the column density determination.

As seen in Fig.~\ref{fig:CSlvg} shows, the best fit contour does not cover the $J=2-1$ data at a probability level of 99.5$\%$.
This fact suggests that the CS emission is
best explained if one consider several gas components
in the region covered by the telescope beam. These components are related to the need of two different excitation temperatures to fit the
population diagram.
The $J=2-1$ to $1-0$ ratio of 1.5 derived by \citet{Paglione}, also plotted in Fig~\ref{fig:CSlvg}, supports this scenario.
For the  180\,km\,s$^{-1}$ velocity component (Fig.~\ref{fig:CSlvg}(a)) a higher density component (n(H$_2$)$\sim4\times10^5$ cm$^{-3}$)
explains the $J_{\rm u}=5,4,3$ transitions, and a lower density component (n(H$_2$)$\sim6\times10^4$ cm$^{-3}$), with a 3 times higher
column density, explains the lower $J$ transitions. Similar results, within the errors, can be derived for the 280\,km\,s$^{-1}$ 
component.

From the non-LTE calculations we derive C$^{32}$S/C$^{34}$S ratios of 4--9 and $^{12}$CS/$^{13}$CS ratios of 22--35 for the 180 and
280\,km\,s$^{-1}$ velocity components respectively. The $^{12}$CS/$^{13}$CS intensity ratio obtained for the 180\,km\,s$^{-1}$ component
 is half the $^{12}$C/$^{13}$C ratio estimated by \citet{Henkel93} while that obtained for the 280\,km\,s$^{-1}$ component is closer to
their value. This may be explained by the fact that the $^{13}$CS observation by \citet{Henkel93} where made towards a position $\sim6''$
closer to the emission peak of the 280\,km\,s$^{-1}$ component.

\section{Discussion}
\subsection{Molecular fractional abundances}
We find that most of the sulfur-bearing species detected towards the nuclear region of NGC\,253 show similar fractional
 abundances of a few 10$^{-9}$ (Table~\ref{tab:MolecDensity}).
Only CS and OCS have abundances a factor of $\sim5$ higher.
In order to understand the mechanism that dominates the chemistry in the nuclear environment of NGC\,253 we
summarize in Table~\ref{tab:comparison} the fractional abundances of all the sulfur-bearing molecules detected towards this
starburst galaxy and those observed towards selected Galactic sources
(hot cores, Galactic center clouds, dark clouds, photodissociation regions (PDRs) and shocks driven by massive protostars)
which are dominated
by different kinds of chemistry. The abundance ratios discussed on this section are also shown in Table~\ref{tab:comparison}.

The SO$_2$ abundance in PDRs is much lower,by a factor of 40, than that found in NGC\,253.
Thus, abundance ratios such as H$_2$S/SO$_2$ and CS/SO$_2$ can be used to establish the role of photodissociation
in the chemistry of NGC\,253. Particularly instructive is the SO/SO$_2$ abundance ratio. SO/SO$_2$ $\sim100$ towards PDRs is about
two orders of magnitude larger than that in NGC\,253 which suggests that photodissociation does not play a major role in the chemistry
of its nuclear region.

A comparison with dark clouds shows a clearly lower abundance of OCS than that
observed towards NGC\,253. The OCS/SO ratio is 0.1 and 0.4 in L134N and TMC-1
respectively, while we measure $\sim 4$ in NGC\,253. The abundance ratios of SO$_2$/SO and H$_2$S/SO of $\sim 0.2$ and $\sim 0.05$
respectively measured towards these dark clouds are also much smaller than the corresponding ratios of $\sim 1$ observed towards NGC\,253.
Therefore, dark molecular
clouds like L134N and TMC-1 do not seem to dominate the molecular composition near the nucleus of NGC\,253.

Chemistry driven by high velocity shocks in molecular outflows
also differs significantly from the chemistry found in NGC\,253.
The OCS/SO ratio measured towards these sources
is two orders of magnitude smaller than that in NGC\,253, while CS/H$_2$S is three orders of magnitude lower.
The large abundance of H$_2$S, as well as the low abundance of molecules such as CS, OCS, and H$_2$CS in local molecular
outflows, is not observed in NGC\,253.

As far as the hot cores are concerned, the molecular abundances vary significantly between Galactic sources. Different evolutionary states
may account for these observed differences. As seen in Table~\ref{tab:comparison} the abundance ratios in the
Orion hot core mostly disagree with those in NGC\,253. The relative abundances of most of the species towards SGR\,B2(N) are similar
to what we find in NGC\,253 except for the underabundance of H$_2$S and OCS with respect to the other molecules.
The ratio OCS/SO is almost two orders of magnitude lower in both hot cores than in NGC\,253.
In addition, the low rotational temperatures ($T_{\rm rot}<25$K) derived for all the sulfur-bearing species in NGC\,253 suggest that the observed
emission cannot be due to hot cores where much higher rotational temperatures are expected ($T_{\rm kin}>70$K).

\citet{Martin03} pointed out that the molecular abundances of SO$_2$, NO, and NS are similar to those
found in the envelope of the SGR\,B2 molecular complex.
Table~\ref{tab:comparison} also includes the comparison with two different clouds
in the envelope of SGR\,B2.
The first position, labeled SGR\,B2(OH), is located $30''$ south of SGR\,B2\,M, and
the second one $100''$ north-east of SGR\,B2\,M.
SO$_2$, SO, and H$_2$CS abundances in the Galactic center clouds agree well with those observed towards NGC\,253.
OCS/SO abundances ratios of 5 and 9 are similar to the ratio of $\sim 4$ found in NGC\,253.

Our study of the sulfur chemistry towards the nuclear region of NGC\,253 seems to indicate a chemistry similar to that of the envelope of
SGR\,B2 where low velocity shocks are thought to be the main heating mechanism of the molecular material
\citep{Flower,Pintado97,Pintado01}.
The low rotational temperatures derived for NGC\,253 are also in agreement with those derived by \citet{Cummins} in the SGR\,B2 envelope.
These results support the idea of large scale low velocity shocks driving the chemistry of the inner molecular material
in the nucleus of NGC\,253 which is consistent with the high rotational temperatures derived from NH$_3$ and the large observed 
SiO abundances \citep{Burillo00,Mauers03}.

\subsection{Chemical models}
In order to get an insight into the dominant chemistry around the central region of NGC\,253,
we compare our results with those available from chemical models \citep{Millar97,Charnley97,Hatchell98,Buckle03}.
For this we will assume that the whole nuclear region of NGC\,253 can be described as a single molecular cloud.
This is not likely as we are observing a 200\,pc region where a number
of unresolved molecular complexes with different physical and chemical properties coexist.
Nevertheless, the dominant observed chemistry in the nuclear region of NGC\,253 could be characterized by a single giant cloud as observed
for the envelope of SGR\,B2 which has a size of $\sim30$\,pc. The SiO mapping of the Galactic center region show molecular cloud complexes
extended over even larger scales \citep{Pintado97}.

We have compared the observed molecular abundances in NGC\,253 with the time-dependent models used by \citet{Hatchell98}
and  \citet{Buckle03} to describe the chemistry of hot cores and low mass protostars respectively.
These models assume the release of H$_2$S from grain mantles as the main precursor of the sulfur chemistry.
Models with $T_{\rm kin}\leq 50\,\rm K$ and n(H$_2$)$<10^5{\rm cm}^{-3}$
seem to best reproduce the abundances measured for SO$_2$, SO, H$_2$S, and H$_2$CS shown in Table~\ref{tab:comparison}.
This result is in agreement with the idea that the emission from the nuclear region of NGC\,253 is dominated by low excitation temperature
and moderately dense molecular clouds similar to those found in the Galactic center clouds.

However current chemical models fall short of accounting for the observed abundances of OCS by more than one order of magnitude.
It has been pointed out by \citet{Hatchell98} and \citet{Vandertak03}
that OCS may play an important role in the chemistry of sulfur-bearing molecules.
Near-IR observations of massive protostars \citep{Geballe95,Tielens89,Palumbo95}
 show that sulfur may freeze out onto grains not only in the form of H$_2$S but also in the form of OCS.
In fact OCS is the only sulfur bearing molecule detected so far in icy grain mantles.
This would imply that OCS is also sputtered from grains by shocks.
\citet{Hatchell98} revised the chemical model by including an initial large abundance of OCS as well
as reducing that of H$_2$S, assuming these species are initially injected into the gas phase.
These models, depending on the initially injected H$_2$S and OCS \citep{Wakelam04},
may account for the high OCS abundance observed in the nuclear region of NGC\,253.

\subsection{Sulfur isotopic ratios}
The $^{32}$S/$^{34}$S abundance ratios of $8\pm2$ and $13.5\pm2.5$ observed towards the nuclear starburst environment of NGC\,253
(Sect.~\ref{sect:results}) and NGC\,4945 \citep{Wang04}
respectively are similar to those found within the Galactic inner 3\,kpc region, but considerably lower than the value
of $\sim 24$ measured in the Galactic disk \citep{Chin}.
The estimated lower limit of the $^{34}$S/$^{33}$S ratio towards NGC\,253 is higher
than the abundance of $\sim6$ measured towards most of the Galactic sources studied by \citet{Chin}.

Massive stars, as well as type Ib/c and II supernovae, appear to slightly overproduce  $^{34}$S and underproduce $^{33}$S compared to
$^{32}$S \citep[see][and references therein]{Chin}.
Thus, the compact stellar clusters observed near the center of NGC\,253 \citep{Watson96} and the estimated overall star formation 
rate of 3.6\,M$_\odot$yr$^{-1}$ \citep{Strickland04} and supernova rate (SNr) of 0.05--0.3\,yr$^{-1}$
\citep{Mattila01,Ulvestad97} might account for the low $^{32}$S/$^{34}$S as well as
the enhanced $^{34}$S/$^{33}$S ratios in NGC\,253. 
This fact is supported by the high massive star formation rate of $\sim 0.1$\,M$_\odot$yr$^{-1}$ estimated by \citet{Forbes93} for the
inner $6\arcsec$ of this starburst.

Uncertainties are large enough that the $^{32}$S/$^{34}$S ratio observed towards NGC\,253 may not significantly differ from that in NGC\,4945,
where slightly higher star formation and supernova rates \citep{Strickland04} might also have induced an overproduction of $^{34}$S in its nuclear
region.

\acknowledgements
J.\,M.-P. has been partially supported by the Ministerio de Ciencia Y Tecnolog\'{\i}a with grant ESP2002-01627, AYA2002-10113E, AYA2003-0090
and ESP2004-00665. We wish to thank F.F.S. van der Tak for critical reading of the manuscript.

\newpage

\clearpage

\newpage

\begin{table}[ht]
\caption{Parameters derived from Gaussian fits to the observed features.\label{tab:fits}}
\begin{tabular}{l l c r r r r r}
\tableline
\tableline
Molecule     &$\nu$        & Transition  & $\int{T_{\rm MB}{\rm d}v}$ & $v_{\rm LSR}$             &  $\Delta v_{1/2}$               & $T_{\rm MB}$ & rms\tablenotemark{b} \\
             &(MHz)        &  $J-J'$     &  (mK\,km\,s$^{-1}$)        & (km\,s$^{-1}$)            &  (km\,s$^{-1}$)                 & (mK)         & (mK)                 \\
\tableline
H$_2$S       & 168762.8    &$1_{1,0}-1_{0,1}$&     1520 (180)         &  180                      &     100                         & 14.3         & 4.7  \\
             &             &             &         2250 ( 80)         &  275                      &      88                         & 23.9         &      \\
CS           &  97981.0    & $2-1$       &    4400 (500)              &  172                      &      96                         & 43.4         & 5.2  \\
             &             &             &    8200 (500)              &  290                      &     116                         & 66.5         &      \\
             & 146969.0    & $3-2$       &   11900 (200)              &  185                      &     100                         & 111.2        & 4.1  \\
             &             &             &   13700 (200)              &  288                      &     117                         & 110.3        &      \\
             & 195954.2    & $4-3$       &   11500 (400)              &  185\tablenotemark{a}     &     100                         & 108.2        & 7.0  \\
             &             &             &   13400 (500)              &  288\tablenotemark{a}     &     121                         & 104.5        &      \\
             & 244935.6    & $5-4$       &    4400 (400)              &  158                      &     107                         & 38.2         & 10.8 \\
             &             &             &    5400 (130)              &  262                      &     108                         & 46.7         &      \\
C$^{34}$S    & 144617.1    & $3-2$       &    3000 (300)              &  188                      &     105                         & 27.1         & 1.4  \\
             &             &             &    1500 (300)              &  284                      &      86                         & 16.1         &      \\
$^{13}$CS    & 138739.3    & $3-2$       &    550  ( 70)              &  188                      &      70                         &  7.4         & 1.9  \\
             &             &             &    390  ( 50)              &  274                      &      70\tablenotemark{a}        &  5.2         &      \\
C$^{33}$S    & 145755.8    & $3-2$       &    $<600 $                 &...                        &...                              & $<3.5$       & 2.8  \\
SO           & 138178.6    &  $4_3-3_2$  &    1400 (200)              &  180                      &      73                         &  17.8        & 2.1  \\
             &             &             &    1100 (200)              &  274                      &      88                         &  11.6        &      \\
             & 158971.8    &  $3_4-2_3$  &     700 (200)              &  180\tablenotemark{a}     &      92                         &   7.2        & 2.9  \\
             &             &             &    1000 (200)              &  274\tablenotemark{a}     &      84                         &  11.4        &      \\
H$_2$CS      & 135297.8    &$4_{1,4}-3_{1,3}$&     1040 ( 90)         &  221                      &     158                         &  6.2         & 1.4  \\
             & 139483.4    &$4_{1,3}-3_{1,2}$&     1340 (150)         &  221                      &     176                         &  7.2         & 2.0  \\
OCS          &  85139.1    & $7-6$       &    1100 (300)              &  212\tablenotemark{a}     &     210\tablenotemark{a}        &  5.1         & 2.7  \\
             &  97301.2    & $8-7$       &    1080 (120)              &  214                      &     219                         &  4.6         & 1.8  \\
             & 133785.9    & $11-10$     &    1220 (130)              &  212                      &     200                         &  5.8         & 1.3  \\
             & 145946.8    & $12-11$     &     580 (170)              &  212\tablenotemark{a}     &     210\tablenotemark{a}        &  2.6         & 1.4  \\
\tableline
\end{tabular}
\tablenotetext{a}{fixed parameter.}
\tablenotetext{b}{rms values in $T_{\rm mb}$ and smoothed to a 10\,km\,s$^{-1}$ channel width.}
\end{table}

\begin{table}[ht]
\caption{Physical parameters derived for each species.\label{tab:MolecDensity}}
\begin{tabular}{l l l l l l}
\tableline
\tableline
\multicolumn{2}{c}{Molecule} & $N$            & $T_{\rm rot}$  & [X]/[H$_2$]\tablenotemark{a}              \\
		&	     &(10$^{13}$cm$^{-2}$) & (K)            &  10$^{-9}$                        \\
\tableline
H$_2$S		&(180\,km/s) & 2.4            & 12   		  & 1.4           		\\
		&(270\,km/s) & 3.6            & 12   		  & 2.1           		\\
CS\tablenotemark{b} &(180\,km/s) & 20         &  9.7(0.4)	  & 12            		\\
   		&(280\,km/s) & 14             &  10 (0.2)	  & 8.2           		\\
NS\tablenotemark{c}&	     & 5              & 8(1)		  & 3               \\
SO		&(180\,km/s) &   3            & 23(14)	  	  & 2           		\\
		&(270\,km/s) & 2.5            & 23    	  	  & 1.5           		\\
H$_2$CS		&	     & 4.6            & 12    		  & 2.7           		\\
OCS		&	     & 26             & 16(2)             & 15 \\
SO$_2$\tablenotemark{c}&     & 7              & 14(9)		  & 4               \\
\tableline
\end{tabular}
\tablecomments{Column densities are source averaged values over a 20\arcsec region.}
\tablenotetext{a}{Assumed $N({\rm H}_2)=1.7\,10^{22}{\rm cm}^{-2}$ from \citet{Mauers03}.}
\tablenotetext{b}{Column density derived using the optically thin $^{13}$CS line and assuming a ratio $^{12}$C/$^{13}$C=40 \citep{Henkel93}.}
\tablenotetext{c}{from \citet{Martin03}.}
\end{table}

\begin{table}[ht]

\caption{Relative abundances of S-bearing molecules in different environments. \label{tab:comparison}}
\begin{tabular}{l   c   c   c   c  c   c   c   c   c  c}
\tableline
\tableline
          &                       &\multicolumn{2}{c}{Hot Cores}                    &\multicolumn{2}{c}{Galactic center clouds}               &\multicolumn{2}{c}{Dark clouds}            &PDRs                    &Outflows                    \\
          & NGC                   &SGR\,B2                &Orion                    &SGR\,B2                  &Sgr B2                         &L134N                &TMC-1                &Orion                   &Orion                       \\
Molecule  & 253                   &(N)\tablenotemark{(a)} &K-L\tablenotemark{(b)}   &(OH)\tablenotemark{(c)}  &(envelope)\tablenotemark{(d)}  &\tablenotemark{(e)}  &\tablenotemark{(e)}  &Bar\tablenotemark{(f)}  &Plateau\tablenotemark{(b)}  \\
\tableline
H$_2$S    &4                      &0.2\tablenotemark{(g)} &5000\tablenotemark{(h)}  &...                      &1.5                            &0.8                   &$<$0.5              &6                       &4000\tablenotemark{(h)}     \\
CS        &20                     &...                    &6                        &...                      &11                             &1                     &10                  &20                      &4                           \\
NS        &3\tablenotemark{(i)}   &10                     &0.4\tablenotemark{(j)}   &...                      &2.1                            &0.4\tablenotemark{(k)}&1\tablenotemark{(k)}&...                     &...                         \\
SO        &4                      &20                     &190                      &4.4                      &6                              &20                    &5                   &9                       &200                         \\
H$_2$CS   &3                      &20                     &0.8                      &6                        &5                              &0.6                   &3                   &...                     &8                           \\
OCS       &15                     &2                      &11                       &21                       &53                             &2                     &2                   &...                     &10                          \\
SO$_2$    &4\tablenotemark{(i)}   &30                     &120                      &5.8                      &5.4                            &4                     &$<$1                &0.1                     &100                         \\
\tableline
H$_2$S/SO & 1                     & 0.01                  & 26                      & ...                     & 0.25                          & 0.04                 & $<$0.1             &0.7                     &20                          \\
CS/SO$_2$ & 5                     & ...                   & 0.005                   & ...                     & 2                             & 0.25                 & $<$10              & 200                    &0.04                        \\
SO$_2$/SO & 1                     & 1.5                   & 0.6                     & 1.3                     & 0.9                           & 0.2                  & $<$0.2             & 0.01                   & 0.5                        \\
OCS/SO    & 3.75                  & 0.1                   & 0.06                    & 4.8                     & 8.8                           & 0.1                  & 0.4                &...                     & 0.05                       \\
CS/H$_2$S & 5                     & ...                   & 0.001                   & ...                     & 7.3                           & 1.25                 & $>$20              & 3.3                    & 0.001                      \\
\tableline
\end{tabular}
\tablecomments{Units of 10$^{-9}$ are used.}
\tablerefs{
${(a)}$ \citet{Nummelin};
${(b)}$ \citet{Sutton95};
${(c)}$ \citet{Cummins} assuming N(H$_2$)=$1\times10^{23}{\rm cm}^{-2}$;
${(d)}$ \citet{MartinPrep};
${(e)}$ \citet{Ohishi};
${(f)}$ \citet{Jansen95};
${(g)}$ \citet{Minh91};
${(h)}$ \citet{Minh90};
${(i)}$ \citet{Martin03};
${(j)}$ \citet{McGonagle97};
${(k)}$ \citet{McGonagle94};
}
\end{table}


\newpage

\begin{thebibliography}{}
\bibitem[Buckle \& Fuller(2003)]{Buckle03} Buckle, J.V., \& Fuller, G.A. 2003, \aap, 399, 567
\bibitem[Charnley(1997)]{Charnley97} Charnley, S.B. 1997, \apj, 481, 396
\bibitem[Chin et al.(1996)]{Chin} Chin, Y.-N., Henkel, C., Whiteoak, J.B., Langer, N., \& Churchwell, E.B. 1996, \aap, 305, 960
\bibitem[Cummins et al.(1986)]{Cummins} Cummins, S.E., Linke R.A., \& Thaddeus P. 1986, \apjs, 60, 819
\bibitem[Flower et al.(1995)]{Flower} Flower, D.R., Pineau des For\^ets, G., \& Walmsley, C.M. 1995, \aap, 294, 815
\bibitem[Forbes et al.(1993)]{Forbes93} Forbes, D.A., Ward, M.J., Rotaciuc, V., Blietz, M., Genzel, R., Drapatz, S., van der Werf, P.P., \& Krabbe, A. 1993, ApJ, 406, L11
\bibitem[Geballe et al.(1985)]{Geballe95} Geballe, T.R., Baas, F., Greenberg, J.M., \& Schutte, W. 1985, \aap, 146, L6
\bibitem[Garc\'{\i}a-Burillo et al.(2000)]{Burillo00} Garc\'{\i}a-Burillo, S., Mart\'{\i}n-Pintado, J., Fuente, A., \& Neri, R. 2000, \aap, 355, 499
\bibitem[Hatchell et al.(1998)]{Hatchell98} Hatchell, J., Thompson, M.A., Millar, T.J., \& Macdonald, G.H. 1998, \aap, 338, 713
\bibitem[Hatchell \& Viti(2002)]{Hatchell02} Hatchell, J., \& Viti, S. 2002, \aap, 381, L33
\bibitem[Henkel \& Bally(1985)]{Henkel85} Henkel, C., \& Bally, J. 1985, \aap, 150, L25
\bibitem[Henkel et al.(1993)]{Henkel93} Henkel, C., Mauersberger, R., Wiklind, T., H\"uttemeister, S., Lemme, C., \& Millar, T.J. 1993, \aap, 268, L17
\bibitem[Heikkil\"a et al.(1999)]{Heikkila99} Heikkil\"a, A., Johansson, L. E. B., \& Olofsson, H. 1999, \aap, 344, 817
\bibitem[Jansen et al.(1995)]{Jansen95} Jansen, D.J., Spaans, M., Hogerheijde, M.R., \& van Dishoeck, E.F. 1995, \aap, 303, 541
\bibitem[Mart\'{\i}n et al.(2003)]{Martin03} Mart\'{\i}n, S., Mauersberger, R., Mart\'{\i}n-Pintado, J., Garc\'{\i}a-Burillo, S., \& Henkel, C.  2003, \aap, 411, L465
\bibitem[Mart\'{\i}n et al.(in preparation)]{MartinPrep} Mart\'{\i}n, S., Mart\'{\i}n-Pintado, J., Mauersberger, R., Requena, M.A., in preparation
\bibitem[Mart\'{\i}n-Pintado et al.(1997)]{Pintado97} Mart\'{\i}n-Pintado, J., de Vicente, P., Fuente, A., \& Planesas, P. 1997, \apj, 482, L45
\bibitem[Mart\'{\i}n-Pintado et al.(2001)]{Pintado01} Mart\'{\i}n-Pintado, J., Rizzo, J.R., de Vicente, P., Rodr\'{\i}guez-Fern\'andez, N.J., Fuente, A. 2001, \apj, 548, L65
\bibitem[Mattila \& Meikle(2001)]{Mattila01} Mattila, S. \& Meikle, W.P.S. 2001, \mnras, 324, 325
\bibitem[Mauersberger \& Henkel(1989)]{Mauers89} Mauersberger, R., \& Henkel, C. 1989, \aap, 223, 79
\bibitem[Mauersberger et al.(1995)]{Mauers95} Mauersberger, R., Henkel, C., \& Chin, Y.-N., 1995, \aap, 294, 23
\bibitem[Mauersberger et al.(1996)]{Mauers96} Mauersberger, R., Henkel, C., Wielebinski, R., Wiklind, T., \& Reuter, H.-P. 1996, \aap, 305, 421
\bibitem[Mauersberger et al.(2003)]{Mauers03} Mauersberger, R., Henkel, C., Wei\ss, A., Peck, A.B., \& Hagiwara, Y. 2003, \aap, 403, 561
\bibitem[McGonagle et al.(1994)]{McGonagle94} McGonagle, D., Irvine, W.M., \& Ohishi, M. 1994, \apj, 422, 621
\bibitem[McGonagle \& Irvine(1997)]{McGonagle97} McGonagle, D., \& Irvine, W.M. 1997, \apj, 477, 711
\bibitem[Millar et al.(1997)]{Millar97} Millar, T.J., Macdonald, G.H., \& Gibb, A.G. 1997, \aap, 325, 1163
\bibitem[Minh et al.(1990)]{Minh90} Minh, Y.C., Irvine, W.M., McGonagle, D., \& Ziurys, L.M. 1990, \apj, 360, 136
\bibitem[Minh et al.(1991)]{Minh91} Minh, Y.C., Ziurys, L.M., Irvine, W.M., \& McGonagle, D. 1991, \apj, 366, 192
\bibitem[Nummelin et al.(2000)]{Nummelin} Nummelin A.,  Bergman P., Hjalmarson \AA., Friberg, P., Irvine, W. M., Millar, T. J., Ohishi, M., \& Saito, S. 2000, \apjs, 128, 213
\bibitem[Ohishi et al.(1992)]{Ohishi} Ohishi, M., Irvine, W. M., \& Kaifu, N. 1992, in IAU Symp. 150, Astrochemistry of Cosmic Phenomena, ed. P.D. Singh (Dordrecht: Kluwer), 171
\bibitem[Paglione et al.(1995)]{Paglione} Paglione, T. A. D., Jackson, J. M., Ishizuki, S., \& Nguyen-Q-Rieu 1995, \aj, 109, 1716
\bibitem[Palumbo et al.(1995)]{Palumbo95} Palumbo, M.E., Tielens, A.G.G.M., \& Tokunaga, A.T. 1995, \apj, 449, 674
\bibitem[Peng et al.(1996)]{Peng96} Peng, R., Zhou, S., Whiteoak, J. B., Lo, K. Y., \& Sutton, E. C. 1996, \apj, 470, 821
\bibitem[Petuchowski \& Bennet(1992)]{Petu92} Petuchowski, S. J., \& Bennett, C. L. 1992, \apj, 391, 137
\bibitem[Strickland et al.(2004)]{Strickland04} Strickland, D. K., Heckman, T. M., Colbert, E. J. M., Hoopes, C. G., \& Weaver, K. A. 2004, \apj, 606, 829
\bibitem[Sutton et al.(1995)]{Sutton95} Sutton, E.C., Peng, R., Danchi, W.C., Jaminet, P.A., Sandell, G., \& Russell, A.P.G. 1995, \apjs, 97, 455
\bibitem[Tielens(1989)]{Tielens89} Tielens, A.G.G.M. 1989, in IAU Symp. 135, Interstellar Dust, ed. L.J. Allamandola \& A.G.G.M. Tielens (Dordrecht: Kluwer), 239
\bibitem[Ulvestad \& Antonucci(1997)]{Ulvestad97} Ulvestad, J.S \& Antonucci, R.R.J. 1997, \apj, 488, 621
\bibitem[van der Tak et al.(2003)]{Vandertak03} van der Tak, F.F.S., Boonman,A.M.S., Braakman, R., \& van Dishoeck, E.F. 2003, \aap, 412, 133
\bibitem[Wakelam et al.(2004)]{Wakelam04} Wakelam, V., Caselli, P., Ceccarelli, C., Herbst, E. \& Castets, A. 2004, \aap, 422, 159
\bibitem[Wang et al.(2004)]{Wang04} Wang, M., Henkel, C., Chin, Y.-N., Whiteoak, J., Cunningham, M., Mauersberger, R., \& Muders, D. 2004, \aap, 422, 883
\bibitem[Watson et al.(1996)]{Watson96} Watson, A. M., et al. 1996, \aj, 112, 534
\end{thebibliography}
\end{document}